\documentstyle[prl,aps]{revtex}
\begin{document}

\input{epsf.tex}
\epsfverbosetrue

\bibliographystyle{unsrt} 

\title{Trapped Bose-Einstein Condensates: Role of Dimensionality}

\author{Yuri S. Kivshar and Tristram J. Alexander}

\address{Optical Sciences Center, Research School of Physical
Sciences and Engineering\\ Australian National University,
Canberra ACT 0200, Australia}

\maketitle

\begin{abstract}
We analyse systematically, from the viewpoint of the nonlinear physics 
of solitary waves, the effect of the spatial dimension ($D = 1,2,3$) 
on the structure and stability of the Bose-Einstein condensates (BECs) trapped in 
an external anisotropic parabolic potential.  While for the positive scattering length 
the stationary ground-state solutions of the Gross-Pitaevskii equation are 
shown to be always stable independently of the spatial dimension, for the negative 
scattering length the ground-state condensate is stable only in the 1D and 
2D cases, whereas 
in the 3D case it becomes unstable.  A direct link between nonlinear modes of BECs 
and (bright and dark) solitary waves of the 
nonlinear Schr\"{o}dinger equation is demonstrated for all the dimensions.
\end{abstract}

\section{Introduction}

Experimental realisations of Bose-Einstein condensation (BEC) in ultracold 
and dilute atomic gases \cite{exp,exp1} generated extensive theoretical 
studies of the macroscopic dynamics of condensed atomic clouds (see, e.g., 
Ref. 3 and references therein).  In experiments, the BEC atoms are 
trapped in a three-dimensional, generally anisotropic, external potential 
created by a magnetic trap, and their collective dynamics can be described 
by the well-known Gross-Pitaevskii (GP) equation, 

\begin{equation}
\label{GPeq}
i\hbar \frac{\partial \Psi}{\partial t} = - \frac{\hbar^{2}}{2m} 
\nabla^{2}_{D} \Psi + V(\vec{r})\Psi + U_{0}|\Psi|^{2}\Psi,
\end{equation}
where $\Psi(\vec{r},t)$ is the macroscopic wave function of a condensate 
in the $D$-dimensional space, $V(\vec{r})$ is a parabolic trapping potential, 
and the parameter $U_{0} = 4\pi \hbar^{2}a/m$ characterises the 
two-particle interaction proportional to the s-wave scaterring length 
$a$.  When $a>0$, the interaction between the particles in the condensate is 
repulsive, as in most current experiments \cite{exp}, whereas for $a<0$ the 
interaction is attractive \cite{exp1}.

From the viewpoint of the nonlinear dynamics of solitary waves, the case of the 
negative scattering length is the most interesting one.  Indeed, it is well known 
that the solutions of Eq. (\ref{GPeq}) without a parabolic potential 
display 
collapse for both $D=2$ and $D=3$, so that {\em no stable spatially localised 
solutions exist in those dimensions at all}.  A trapping potential allows a 
stabilisation of otherwise 
unstable and collapsing solutions and, as has been already shown for the 3D 
case with the help of different approximate techniques \cite{baym,wadati}, there 
exists a critical number $N_{\rm cr}$ of the total condensate particles 
$N$, defined through the normalisation of the condensate wavefunction as
\begin{equation}
\label{Neq}
N = \int |\Psi|^{2}d^{3}\vec{r},
\end{equation}
such that the condensate is stable for $N<N_{\rm cr}$, and it becomes 
unstable for $N\ge N_{\rm cr}$, 
and then its local density $|\Psi|^{2}$ grows to infinity.  This 
critical value of the particles' number is readily observed in experiments 
\cite{exp1}.

In this paper we analyse systematically, in the framework of the GP equation 
(\ref{GPeq}), the structure and stability of a trapped condensate depending on its dimension 
$D$.  In spite of the fact that the first experimental results were 
obtained for the BEC in a 3D anisotropic trap, the cases of lower 
dimensions 
are also of great importance and, moreover, they have a clear physical 
motivation.  First of all, Eq. (\ref{GPeq}) with $D=1$ appears as the 
fundamental model of BEC in highly anisotropic traps of the axial symmetry 
\begin{equation}
\label{Veq}
V(\vec{r}) = \frac{m\omega^{2}}{2}(r_{\perp}^{2}+\lambda z^{2}), \;\;
r_{\perp} = \sqrt{x^{2}+y^{2}},
\end{equation}
provided $\lambda \equiv 
\omega_{z}^{2}/
\omega_{\perp}^{2} \ll 1$, i.e. for {\em cigar-shaped traps}.  In this case, 
the transverse structure of the condensate is controlled by a trapping 
potential, and it can be described by the eigenmodes of a 
two-dimensional isotropic harmonic oscillator.  Then, after averaging over the 
cross-section, Eq. 
(\ref{GPeq}) appears as a 1D model with renormalised parameters, 
and it describes the longitudinal dynamics of the condensate (see Ref. 
6 and Sec. 2 below).

Secondly, in the case $D=2$, the model (\ref{GPeq}) can be employed to 
describe 
the condensate dynamics in the so-called {\em pancake traps} 
($\lambda \gg 1$), and 
it can be related to recent experiments with quasi-two-dimensional 
condensates or coherent atomic systems \cite{2d,2d_1}.

The paper is organised as follows.  In Sec. 2 we consider the 1D case and 
find numerically the ground-state solutions for the trapped condensate in 
both $a<0$ and $a>0$ cases.  In the former case, the problem is found to be 
remarkably similar to that of the soliton propagation in optical 
fibers in the presence of the dispersion management \cite{turitsyn}.  The 
most interesting case of a 2D trap is considered in Sec. 3, where we discuss 
the necessary conditions for the instability of the radially symmetric 
condensate in a trap, and also analyse the invariants of the stationary 
solutions found numerically.  We point out, for the first time to our 
knowledge, that 2D condensates can be stable, in spite of the possibility of 
collapse in the model described by the 2D GP equation.  Similar results are 
briefly presented in Sec. 4 for the 
3D BECs of radial symmetry, and they provide a rigorous background for the variational 
results obtained earlier 
by different methods \cite{baym}.  In all the cases, we also find the condensate 
structure for the positive scattering length where the instability is 
absent in all the dimensions.

\section{Quasi-One-Dimensional Trap}

\subsection{Model}

To derive the 1D GP equation from Eq. (\ref{GPeq}), we assume that the 
parabolic potential $V(\vec{r})$ describes a cigar-shaped trap.  For the
 potential (\ref{Veq}), this means that $\lambda \ll 1$, and the 
transverse structure of the condensate, being close to a Gaussian in shape, is 
mostly defined by the trapping 
potential.  We are interested in the 
stationary solutions of the form $\Psi(\vec{r},t) = \Psi(\vec{r})
{\rm e}^{-\frac{i\mu}{\hbar}t}$, where $\mu$ has the meaning of the chemical 
potential.  Measuring the spatial variables in the units of the longitudinal 
harmonic oscillator length $a_{ho}=(\hbar/m\omega\sqrt{\lambda})^{1/2}$, and 
the wavefunction, in units of $(\hbar\omega/2U_{0}\sqrt{\lambda})^{1/2}$, we 
obtain the following stationary dimensionless equation:
\begin{equation}
\label{Phieq}
\nabla_{D}\Phi + \mu'\Phi - \left[ \frac{1}{\lambda}(\xi^{2} + \eta^{2}) + 
\zeta^{2}\right]\Phi + \sigma|\Phi|^{2}\Phi = 0,
\end{equation}
where $\mu' = 2\mu\sqrt{\lambda}/\hbar\omega$, $(\xi,\eta,\zeta) = 
(x,y,z)/a_{ho}$, and the sign $\sigma = {\rm sgn}(a) = \pm 1$ in front of the nonlinear 
term is defined by the sign of the s-wave scattering length of two-body 
interaction.

We assume that in Eq. (\ref{GPeq}) the nonlinear interaction is weak relative to the trapping potential force, i.e. 
$U_{0}/m\omega \sim \lambda \ll 1$.  Then, it follows from Eq. (\ref{Phieq}) 
that 
the transverse structure of the condensate is of order of $\lambda$, and the 
condensate has a cigar-like shape.  Therefore, we can look for solutions of 
Eq. (\ref{Phieq}) in the form, 
\begin{equation}
\label{divide}
\Phi(r,\zeta) = \phi(r)\psi(\zeta),
\end{equation}
where $r= \sqrt{\xi^{2}+\eta^{2}}$, and $\phi$ is a solution of the auxiliary 
problem for the 2D 
quantum harmonic oscillator 
\begin{equation}
\nabla_{\perp}\phi + \gamma\phi - \lambda^{-1}r^{2}\phi = 0, 
\end{equation}
which we take in the form of the no-node ground state,
\begin{equation}
\label{phi_eq}
\phi_{0}(r) = C\exp{\left(-\frac{r^{2}}{2\sqrt{\lambda}}\right)}, \;\;
\gamma = \frac{2}{\sqrt{\lambda}},
\end{equation}
$C$ being constant.  To preserve all the information about the 3D condensate in an asymmetric trap 
describing its properties by the 1D GP equation for the longitudinal profile, 
we impose the normalisation,
\begin{equation}
\label{normal}
\int^{\infty}_{-\infty}\phi(r)dxdy = 2\pi\int^{\infty}_{0}\phi(r)rdr = 1
\end{equation}
that yields $C^{2} = 1/\pi\sqrt{\lambda}$.

After substituting the solution (\ref{divide}) with the function 
(\ref{phi_eq}) into Eq. (\ref{Phieq}), dividing by $\phi$ and integrating 
over the transverse cross-section $(\xi,\eta)$ of the cigar-shaped condensate, 
we finally obtain the 
following 1D stationary GP equation (see also the similar results in Ref. 6) 
\begin{equation}
\label{GPeq_1}
\frac{d^{2}\psi}{d\zeta^{2}} + \beta\psi - \zeta^{2}\psi + \sigma|\psi|^{2}
\psi = 0,
\end{equation}
where $\beta=\mu'-\gamma$, and the normalisation (\ref{normal}) for $\phi(r)$ 
is used.

Importantly, the number of the condensate particles $N$ is now defined as 
$N = (\hbar\omega/2U_{0}\sqrt{\lambda})Q$, where 
\begin{equation}
\label{Q_eq}
Q = \int^{\infty}_{-\infty} |\psi|^{2}d\zeta
\end{equation}
is the integral of motion for the normalised nonstationary GP equation.

\subsection{Stationary Ground-State Solutions}

Equation (\ref{GPeq_1}) includes all the terms of the same order, and it 
describes the longitudinal profile of the condensate state in an anisotropic 
trap.  In the linear 
limit, i.e. when formally $\sigma \rightarrow 0$, Eq. (\ref{GPeq_1}) becomes 
the well-known 
equation for a harmonic quantum oscillator.  Its localised solutions exist 
only for $\beta \equiv \beta_{n} = 1+2n \; (n=0,1,2,\ldots)$, and they are defined 
through the Gauss-Hermite polynomials, $\psi_{n}(\zeta) = c_{n}{\rm e}^
{-\zeta^{2}/2}H_{n}(\zeta)$, where 
\begin{equation}
\label{Hermite}
H_{n}(\zeta) = (-1)^{n}{\rm e}^{\zeta^{2}/2}\frac{d^{n}({\rm e}^{-\zeta^{2}/2
})}{d\zeta^{n}}, 
\end{equation}
so that $H_{0} = 1$, $H_{1} = 2\zeta$, etc.

To describe the effect of weak nonlinearity, we can employ a perturbation 
theory based on the expansion of the general solution of Eq. (\ref{GPeq_1}) 
in the infinite set of eigenfunctions defined by Eq. (\ref{Hermite}).  The 
similar approach has been used earlier in the theory of the dispersion-managed 
optical solitons (see, e.g., Ref. 10), and it allows us to 
describe the effect of weak interaction on the condensate structure and 
stability.  These results will be presented elsewhere \cite{turitsyn2}.

In the opposite limit, i.e. when the nonlinear term is much larger than the 
parabolic potential, localised solutions exist only for $\beta < 0$ and
$\sigma = +1$, and they are described by the stationary 
nonlinear Schr\"{o}dinger (NLS) equation.  The one-soliton solution is 
\begin{equation}
\label{NLS_1}
\psi_{s}(\zeta) = \frac{\sqrt{-2\beta}}{\cosh(\zeta\sqrt{-\beta})},
\end{equation}
so that $Q_{s} = 4\sqrt{-\beta}$ coincides with the soliton invariant.

In general, the ground-state solution of Eq. (\ref{GPeq_1}) can be found only 
numerically.  Figures 1(a) and 1(b) present some examples of the numerically 
found solutions of Eq. (\ref{GPeq_1}) for several values of the dimensionless 
parameter $\beta$, for both negative (a) and positive (b) scattering length.  
For $\beta \rightarrow 1$, i.e. in the limit of the 
harmonic oscillator mode, the solution is close to Gaussian for both the 
cases.  When $\beta$ deviates from 1, the solution profile is defined by the 
type of nonlinearity.  For attraction ($\sigma = +1$), the profile 
approaches the sech-type soliton (\ref{NLS_1}), whereas for repulsion 
($\sigma = -1$) the solution flattens, cf. Fig. 1(a) and Fig. 1(b).  

In Fig. 2 we show the dependence of the invariant $Q$ on the parameter 
$\beta$, for both the families of localised solutions, corresponding to two 
different signs of the scattering length.  The dotted line shows the limit 
of the soliton solution 
of the NLS equation without a trapping potential.  In the asymptotic, i.e. 
say for $\beta < -2$, the curve $Q_{\rm s}(\beta)$ coincides with the 
invariant $Q$ for the BEC condensate in a trap.  This means that for such 
localised condensates the effect of the trap is negligible and the condensate 
function becomes localised mostly due to an attractive interparticle interaction.

To discuss the stability of localised solutions, we refer to a number of the corresponding 
problems well analysed in nonlinear optics of guided modes and solitary waves 
\cite{yuri}.  In terms of the so-called Vakhitov-Kolokolov stability criterion, 
the condition $dQ/d(-\beta) > 0$ gives the solution stability for the 
case $\sigma = +1$.  Therefore, all 1D solitons for the attraction case are 
stable.  A detailed discussion of the soliton stability for this case can be 
found in Ref. 9.

Formally, the case $\sigma = -1$ corresponds to the opposite sign of the derivative 
$dQ/d\beta$.  However, this case of the so-called defocusing 
nonlinearity is also well-known in nonlinear optics of guided waves, and it 
has been analysed for different types of the trapping potential.  In 
particular, it is known that the Vakhitov-Kolokolov criterion does not apply 
to this case, and all the family of spatially localised solutions are stable 
\cite{arizona}.  Extension of all those results to the case of a parabolic trap 
is a straightforward technical problem.

\subsection{Higher-Order Modes and Multi-Soliton States}

It is well known that in the linear limit $\sigma \rightarrow 0$, Eq. 
(\ref{GPeq_1}) possesses a {\em discrete set of localised modes} corresponding 
to the different orders of the Gauss-Hermite polynomials.  We demonstrate that 
all those modes can be readily obtained for the nonlinear problem as well, 
giving a continuation of the Gauss-Hermite modes to a nonlinear regime.  
Figure 3 shows 
several examples of those modes for both negative ($\sigma = +1$) and positive 
($\sigma = -1$) scattering length, respectively.  In the limit $\beta 
\rightarrow 1$, those modes are defined by the eigenfunctions of the linear 
harmonic oscillator.  The effect of nonlinearity is different for the 
negative and positive scattering length.  For the negative scattering length 
(attraction), the higher-order modes transform into multi-soliton states 
consisting of a sequence of solitary waves with alternating phases [see 
Figs. 3(a) and 3(b)].  This is further confirmed by the analysis of the 
invariant $Q$ vs. $\beta$, where all the branches of the higher-order modes 
tend asymptotically to the soliton dependences $Q_{N} \sim (N+1)Q_{\rm s}$, where $N$ is the order of 
the mode ($N=0,1,\ldots$).  Examples of the dependences $Q(\beta)$ for 
the first- and second-order modes 
are shown in Fig. 4.  From the physical point of view, the higher-order modes 
in the case of attractive interaction correspond to a balance between 
{\em repulsion} of out-of-phase bright solitons and {\em attraction} of the trapping 
potential.  It is clear that such a balance can be easily destroyed by, for 
example, making the soliton amplitudes different.  However, the analysis of 
the stability of such higher-order multi-soliton modes is beyond the scope 
of this paper.

For the positive scattering length ($\sigma = -1$), the higher-order modes 
transform into a sequence of dark solitons \cite{review2}, so that the first-order 
mode corresponds to a single dark soliton, the second-order mode to a 
pair of dark solitons, etc. [see Figs. 3(c) and 3(d)].  Again, these 
stationary solutions satisfy the force balance condition - repulsion between 
dark solitons is exactly compensated by an attractive force of the trapping 
potential.

\section{Quasi-Two-Dimensional Trap}

\subsection{Model and Stationary Ground-State Solutions}

The case of the two-dimensional (2D) GP equation can be associated with a 
quasi-condensate of atomic hydrogen \cite{2d} or quasi-2D gas of laser 
cooled atoms \cite{2d_1}.  Derivation of the GP equation from the first 
principles of scattering theory in the two-dimensional geometry 
is not trivial, and it has been recently 
shown \cite{quasi_2d} that the correct form of the 2D GP equation 
(\ref{GPeq}) should have the 2D interaction potential $U_{0}$ in the form 
$U_{0} = -(2\pi\hbar^{2}/m)\ln^{-1}(ka)$, where $0<ka\le1$, $a$ is the 
scattering length, and the characteristic wavenumber $k$ can be approximated 
as $k \sim 1/a_{ho}$.  Here, we derive the 2D GP equation directly from its 
3D form of Eq. (\ref{GPeq}), for the case of a pancake trap when $\lambda 
\gg 1$.

In the case of a pancake trap, in the parabolic potential 
(\ref{Veq})  we assume the parameter $\lambda$ is large, i.e. $\lambda \gg 1$. 
Then, the longitudinal profile of the 
condensate is controlled by the parabolic potential 
$ \sim (m\lambda\omega^{2}/2)z^{2}$.  
Measuring the spatial variables in the units of the transverse harmonic 
oscillator length $(\hbar/m\omega)^{1/2}$ and the wavefunction in units 
of $(\hbar\omega/2U_{0})^{1/2}$, from Eq. (\ref{GPeq}) we obtain the 
following stationary equation,
\begin{equation}
\label{phieq_2}
\nabla_{D}\Phi + \mu'\Phi-\left[(\xi^{2}+\eta^{2}) + \lambda\zeta^{2}\right]
\Phi + \sigma|\Phi|^{2}\Phi = 0,
\end{equation}
where $\mu' = 2\mu/\hbar\omega$.  For $\lambda \gg 1$, the longitudinal 
structure of the condensate is squeezed in the $z$-direction and its size in 
this direction is of the order of $\lambda^{-1}$.  Therefore, we can look 
for stationary solutions in the form,
\begin{equation}
\label{divide_2}
\Phi(r,\zeta) = \psi(r)\phi(\zeta),
\end{equation}
where $\psi(r)$ depends on the radial coordinate $r=\sqrt{x^{2}+y^{2}}$, and 
this time the function $\phi(\zeta)$ is a solution of the 1D quantum harmonic 
oscillator,
\begin{equation}
\frac{d^{2}\phi}{d\zeta^{2}} + \gamma\phi - \lambda\zeta^{2}\phi = 0,
\end{equation}
with the Gaussian form,
\begin{equation}
\phi_{0}(\zeta) = C\exp{\left(-\frac{\zeta^{2}\sqrt{\lambda}}{2}\right)},\;\;
\gamma = \sqrt{\lambda}.
\end{equation}
Normalisation condition $\int^{\infty}_{-\infty}\phi_{0}^{2}(\zeta)d\zeta =1$ 
yields $C=\pi^{1/4}\lambda^{1/8}$.

Substituting the solution (\ref{divide_2}) into Eq. (\ref{phieq_2}) and 
averaging over $\zeta$, we obtain the following 2D equation for $\psi(r)$, 
\begin{equation}
\label{GPeq_2}
\left(\frac{d^{2}}{dr^{2}} + \frac{1}{r}\frac{d}{dr}\right)\psi + \beta\psi 
- r^{2}\psi + \sigma|\psi|^{2}\psi = 0,
\end{equation}
where $\sigma = \pm 1$ and $\beta = \mu'-\gamma$.  Equation (\ref{GPeq_2}) 
is the 2D GP equation for the stationary solutions without any angular 
dependence.  In the linear limit, i.e. when we 
neglect the nonlinear term ($\sigma \rightarrow 0$), the solution exists only at 
$\beta=2$, and it can 
be written in the form $\psi_{0}(r)=C_{0}\exp{(-r^{2}/2)}$, where $C_{0}$ 
is defined by normalisation.  In the opposite limit, the localised 
solution exists only for $\sigma=+1$, and it is described by the radially 
symmetric solitary wave of the 2D NLS equation.

We integrate Eq. (\ref{GPeq_2}) numerically and find a family of radially 
symmetric localised solutions $\psi(r)$ for the condensate ground state, for both 
$\sigma = +1$ (attraction) and $\sigma = -1$ (repulsion).  Some results 
are presented in Figs. 5(a,b), for different values of the parameter 
$\beta$.  The structure of the localised solutions is similar to those in 
the 1D case.  However, in the limit of large negative $\beta$, the solution 
transforms into the 2D NLS soliton, known to be a self-similar radially symmetric 
solution corresponding 
to a critical collapse.  This is further illustrated in Fig. 6, where we 
present the dependence of the invariant $Q$ on the parameter $\beta$.  
The 2D NLS soliton exists only for 
$\beta<0$ and $\sigma=+1$, and it corresponds to the fixed value of the 
invariant, $Q_{s} \approx 11.7$, shown as a dotted straight line in Fig. 6.

\subsection{Stability and collapse}

Stability of the 2D condensates in a parabolic trap is an important issue.  
In particular, as was shown by Berg\'{e} \cite{berge} and Tsurumi and Wadati 
\cite{wadati}, the presence of a trapping potential does not remove collapse 
from the 2D GP equation.  This result is very much expected because 
it has been known for a long time that, in the case of the 2D NLS equation 
with a parabolic potential, there exists an exact analytical transformation 
that allows the potential to be removed from the 2D NLS equation 
\cite{manassah}.  This means that for some classes of the input conditions 
of the GP equation, the collapse should be observed even in the presence of 
a trapping potential.  Nevertheless, as follows from our results summarised 
in Fig. 6, for the attractive case ($\sigma = +1$) the derivative 
$dQ/d(-\beta)$ is positive and it does not change the sign.  Moreover, according to the 
collapse conditions derived in Ref. 15, for the radius 
$R$ of the time-dependent 2D condensate the following equation holds 
\begin{equation}
\label{eq_R}
\frac{d^{2}\langle R^{2}\rangle}{d t^{2}} = 8H - 4\langle R^{2} \rangle,
\end{equation}
where $H$ is the system Hamiltonian
\begin{equation}
H = 2\pi\int^{\infty}_{0} \left(\left| \frac{d \psi}{dr} \right|^{2} + r^{2}|\psi|^{2}
-\frac{1}{2}|\psi|^{4}\right)rdr
\end{equation}
and $R$ is defined as 
\begin{equation}
\langle R^{2} \rangle = 2\pi\int^{\infty}_{0} r^{3}|\psi|^{2}dr.
\end{equation}
Equation (\ref{eq_R}) means that the condition $H<0$ is a sufficient condition 
for collapse, because the value $\langle R^{2}\rangle$ surely becomes 
negative for some value of $t$ and, therefore, the wavefunction becomes singular 
and collapses as $\langle R^{2} \rangle$ tends to zero.

When $H$ is zero or positive, the dynamics depend on the applied perturbation, 
so that for a large enough perturbation we can lower the stationary value of the 
Hamiltonian $H$.  In Fig. 7 we present the dependences $H(\beta)$ and $H(Q)$ 
for the family of stationary solutions found numerically for the attraction case 
$\sigma = +1$.  It is clear that the whole family of stationary solutions 
corresponds to $H>0$.  Therefore, the ground-state is expected to be stable everywhere,
but its dynamics should be sensitive to the perturbation amplitude for larger $\beta$ 
since a perturbation can make the Hamiltonian negative.  This statement should be 
further verified by numerical simulations of the 2D GP equation.

\subsection{Higher-Order Modes and Vortices}

Similar to the 1D case discussed above, the higher-order modes of a harmonic 
oscillator can be found in the 2D case too, thus providing an analytical 
continuation of the 2D Gauss-Hermite modes.  The set of such solutions is 
broader because it includes a continuation of a superposition of different 
modes.  For example, the mode $\psi_{00}(r)$ is the fundamental ground-state 
solution, that 
continues the zero-order Gauss-Hermite mode $\sim H_{0}(x)H_{0}(y)$, as 
discussed above.  In general, the modes $\psi_{nm}(x,y)$ have no radial 
symmetry and they include both dipole-like and vortex-like states.  

The 
simplest higher-order radially symmetric mode describes the condensate 
with a single vortex, shown in Fig. 8.  Figures 8(a) and 8(b) show the 
density profile of the condensate with a vortex state, for both 
negative ($\sigma = +1$) and positive ($\sigma = -1$) scattering length, 
respectively.  Figure 8(c) presents the dependence of the invariant 
$Q$ vs. $\beta$ for the vortex state in comparison with the ground-state 
mode.  Because the vortex state is a nonlinear mode that extends the corresponding 
mode of the linear system, it 
approaches the value $\beta = 4$.

\section{Three-Dimensional Trap of Radial Symmetry}

The case of a radially symmetric 3D trap is the most studied in the 
literature, so that the corresponding results are well known, and therefore 
we reproduce some of them just for the completeness of the general picture 
of the stationary states.  Similar results for the stationary states 
can be found, e.g. in Refs. 3 and 17, whereas the BEC stability 
has been recently discussed by Huepe {\em et al}. \cite{huepe}.

Figures 9(a) and 9(b) show the density profiles of the 3D condensate of radial 
symmetry for 
different values of $\beta$, for both negative ($\sigma = +1$) and 
positive ($\sigma = -1$) scattering length.  While in the case of repulsion
($\sigma = -1$), the condensate broadens for larger $\beta$ remaining 
always stable, for the attractive interparticle interaction ($\sigma = +1$), 
the condensate is stable only for $\beta < \beta_{\rm cr}$, where 
$\beta_{\rm cr} = 0.72$ corresponds to the maximum value $Q_{\rm cr} \approx 14.45$ 
[see Fig. 10].  This result is consistent with the Vakhitov-Kolokolov 
stability criterion $dQ/d(-\beta) > 0$.  The critical value $\beta_{\rm cr}$ 
corresponds to a critical value of the particles $N_{\rm cr}$ and it has 
been already observed in experiment \cite{exp1}.  Without the trapping potential i.e. in 
the limit of the 3D NLS equation, all 
localised solutions for $\sigma = +1$ are unstable (shown as a dotted curve in 
Fig. 10).

\section{Conclusions}

We have presented a systematic study of the ground-state and higher-order 
spatially localised solutions of the 
GP equation for the Bose-Einstein condensates in a parabolic trap of 
different dimensions.  While many results for the radially symmetric 
ground-state and vortex modes in 3D are available 
in the literature for the condensates with the positive scattering length 
(i.e., for repulsive interaction of the condensate atoms), little is 
known for the condensates with the negative scattering length, especially for 
the 
2D traps.  In particular, we have presented the results indicating that, 
in spite of the collapse 
condition in the 2D case, the family of 2D localised solutions of the GP 
equation can be stable.  Additionally, we have clarified the meaning of 
higher-order localised modes, i.e. {\em nonlinear Gauss-Hermite eigenmodes}, and 
demonstrated their link to multi-soliton states.
We believe the systematic study of the stationary states in all 
the dimensions carried out here from the viewpoint of the nonlinear physics 
of localised states and solitary waves allows us to close a gap in the 
literature and deepens our knowledge about the condensate properties and 
stability.

\begin{figure}
\setlength{\epsfxsize}{12.0cm}
\centerline{\mbox{\epsffile{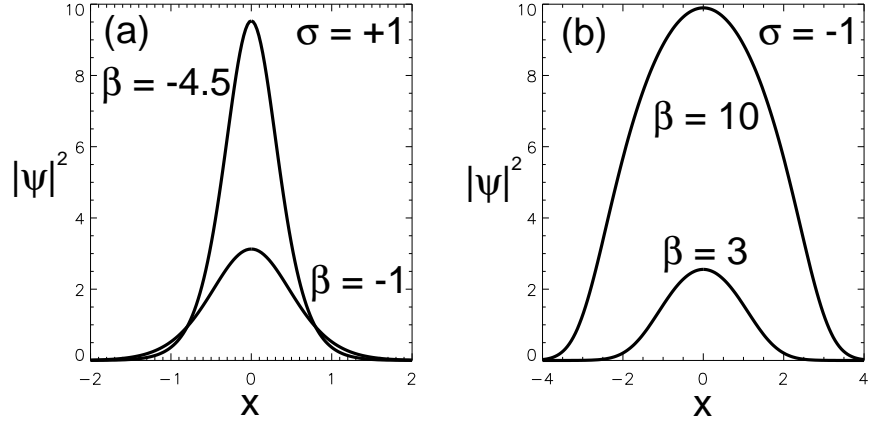}}}
\caption{(a,b)  Examples of the longitudinal condensate density $|\psi|^{2}$ 
described by the solution of Eq. (\ref{GPeq_1}) for the negative 
($\sigma = +1$) and positive ($\sigma = -1$) scattering length, respectively.
  The values of $\beta$ are given next to the curves.}
\label{fig1}
\end{figure}

\begin{figure}
\setlength{\epsfxsize}{12.0cm}
\centerline{\mbox{\epsffile{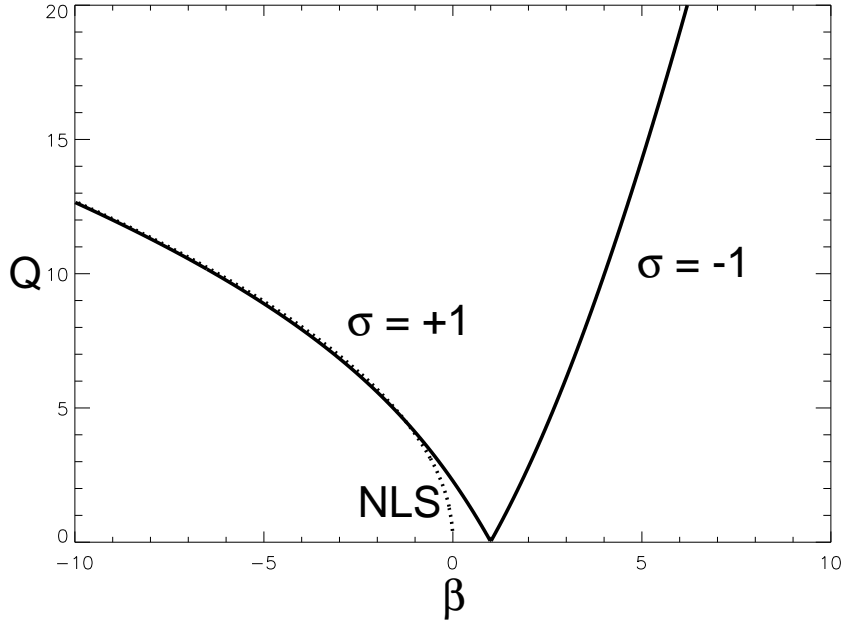}}}
\caption{Dependence of the invariant $Q$, defined by Eq. (\ref{Q_eq}), 
on the parameter $\beta$ for the 
family of the ground-state localised solutions of Eq. (\ref{GPeq_1}).  
Dotted - result for the 1D NLS soliton (\ref{NLS_1}) without a trapping 
potential.}
\label{fig2}
\end{figure}

\begin{figure}
\setlength{\epsfxsize}{12.0cm}
\centerline{\mbox{\epsffile{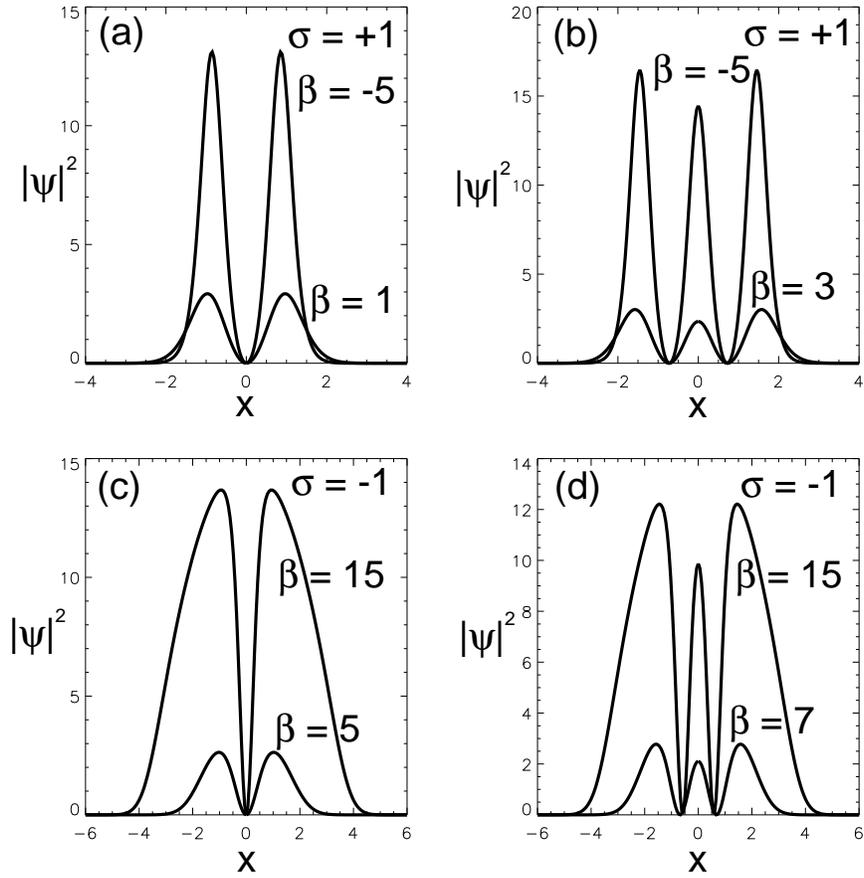}}}
\caption{Examples of higher-order (first and second) nonlinear modes of BEC 
in a parabolic trap shown for the negative ($\sigma = +1$, upper row) and 
positive ($\sigma = -1$, lower row) scattering length.}
\label{fig3}
\end{figure}

\begin{figure}
\setlength{\epsfxsize}{12.0cm}
\centerline{\mbox{\epsffile{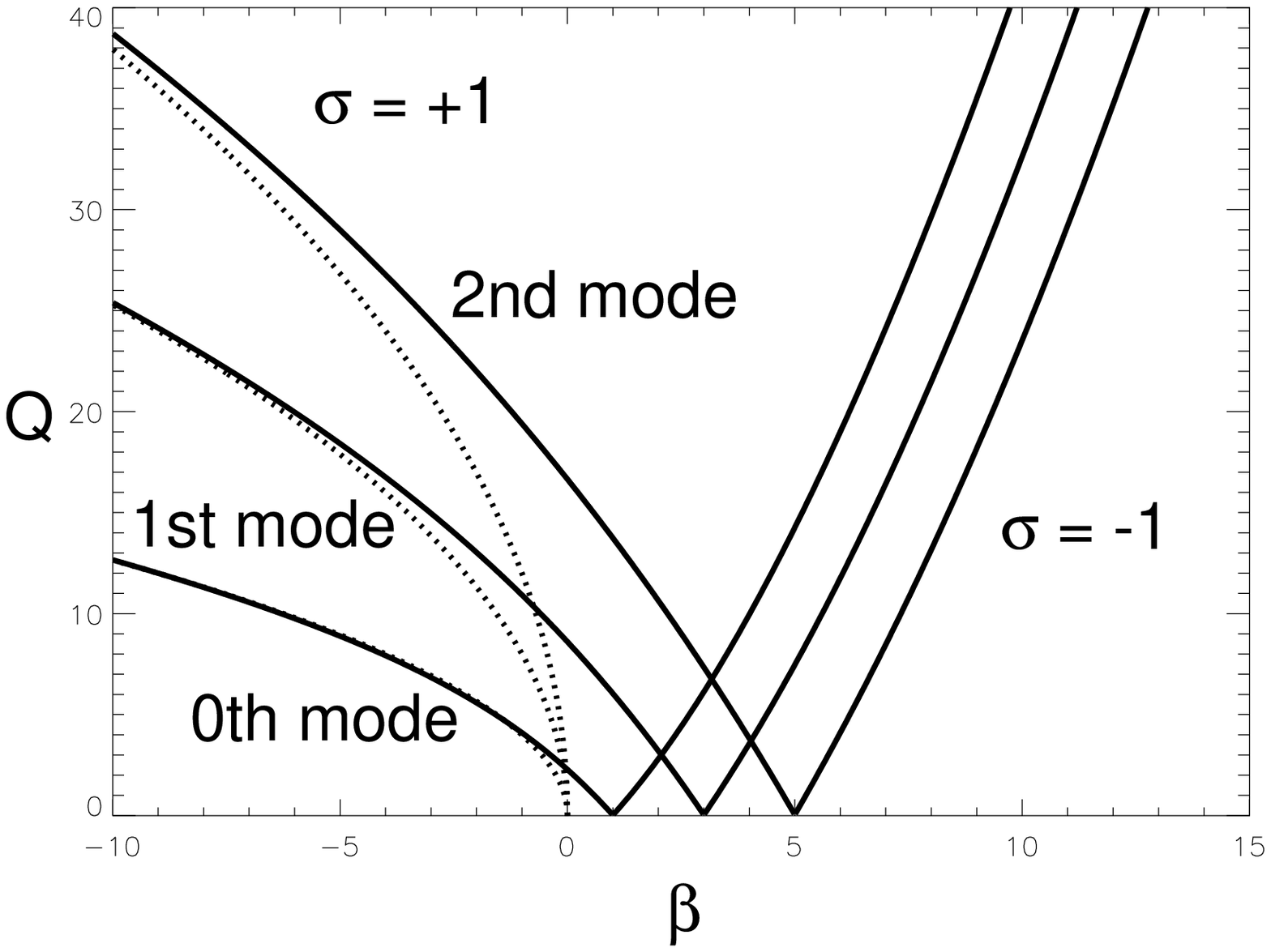}}}
\caption{Invariant $Q$ vs. $\beta$ for the first- and second- order nonlinear modes,
shown together with the corresponding dependence for the fundamental mode of Fig. 2.  
Dotted lines are given by the values $Q_{s}$, $2Q_{s}$, and 
$3Q_{s}$, respectively.}
\label{fig4}
\end{figure}

\begin{figure}
\setlength{\epsfxsize}{12.0cm}
\centerline{\mbox{\epsffile{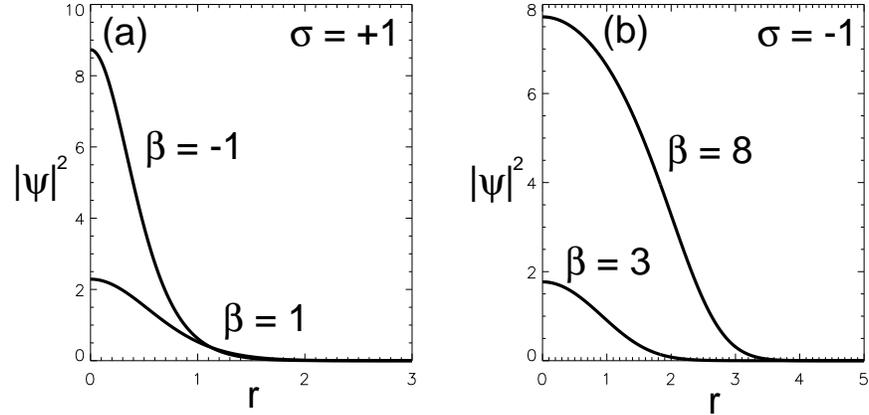}}}
\caption{(a,b) Examples of the transverse condensate density $|\psi|^{2}$ 
described by Eq. (\ref{GPeq_2}) for the attractive ($\sigma = +1$) and 
repulsive ($\sigma = -1$) interaction, respectively.}
\label{fig5}
\end{figure}

\begin{figure}
\setlength{\epsfxsize}{12.0cm}
\centerline{\mbox{\epsffile{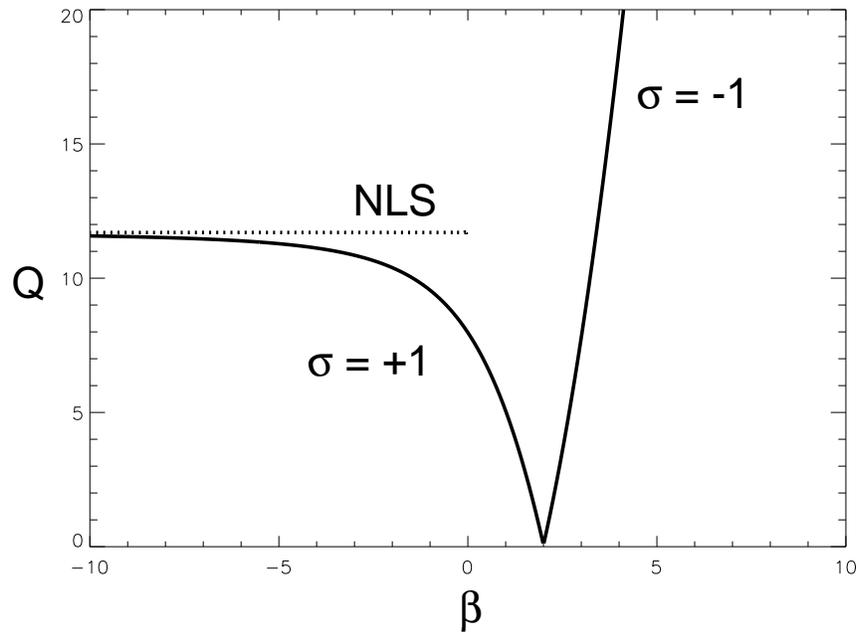}}}
\caption{Dependence of the invariant $Q$ on the parameter $\beta$ for the 
family of the ground-state solutions of Eq. (\ref{GPeq_2}).  Dotted - the 
limit of the 2D NLS soliton without a trapping potential.}
\label{fig6}
\end{figure}

\begin{figure}
\setlength{\epsfxsize}{12.0cm}
\centerline{\mbox{\epsffile{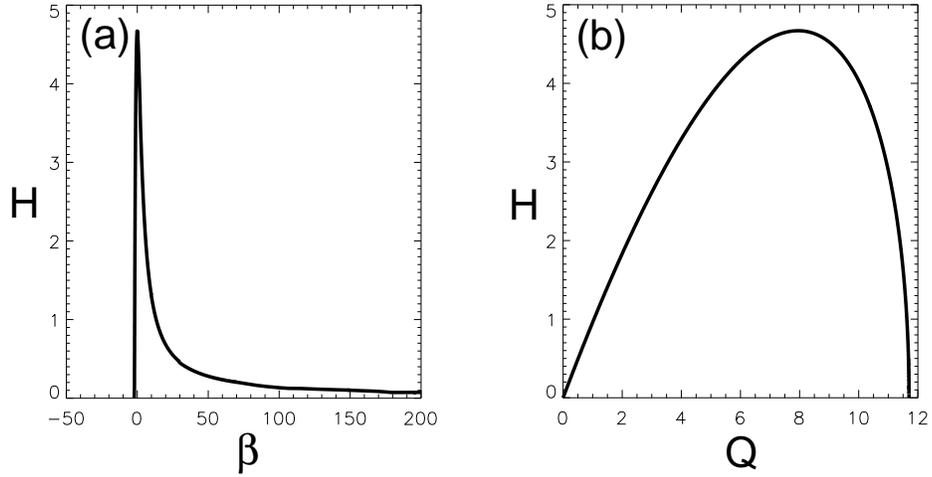}}}
\caption{Dependence of the Hamiltonian $H$ on (a) the parameter $\beta$, and (b) on the 
invariant $Q$ for the family of 2D radially symmetric solutions of Eq. (\ref{GPeq_2}) 
at $\sigma = +1$.}
\label{fig7}
\end{figure}

\begin{figure}
\setlength{\epsfxsize}{12.0cm}
\centerline{\mbox{\epsffile{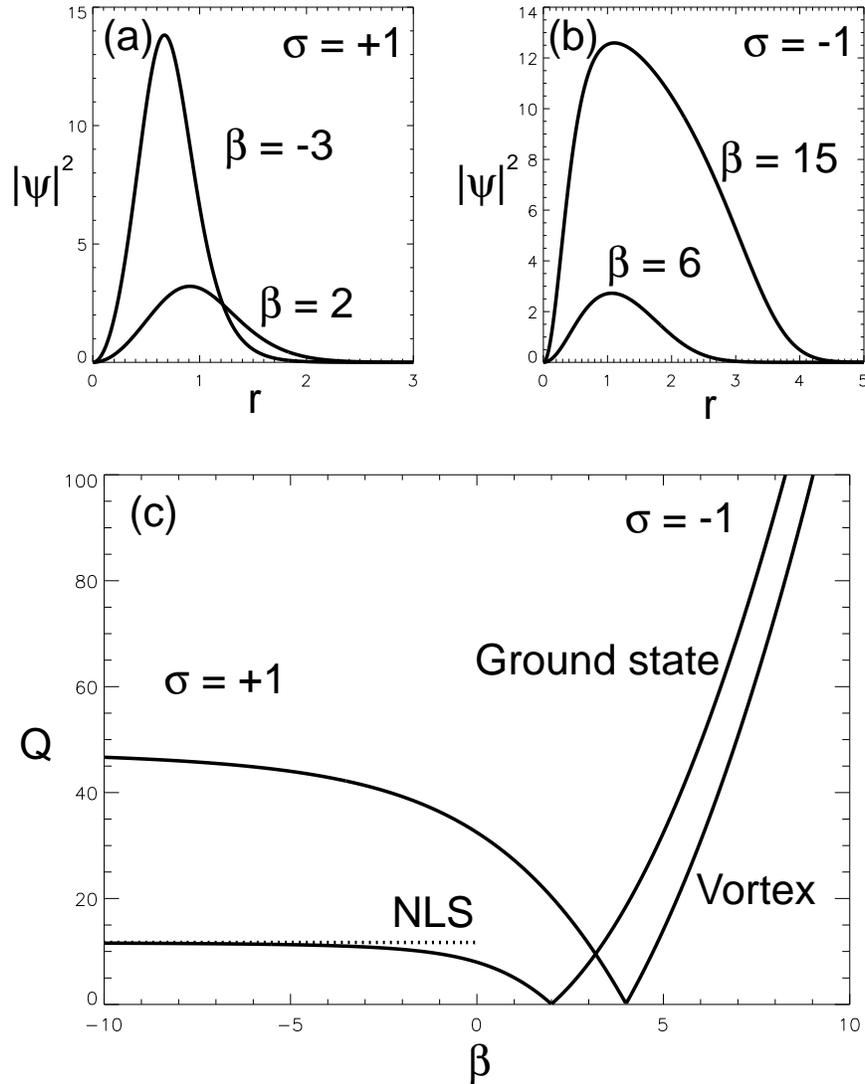}}}
\caption{(a,b) Vortex profiles and (c) the corresponding invariant $Q$ for the ground-state 
and vortex localised 
modes of the 2D GP equation.}
\label{fig8}
\end{figure}

\begin{figure}
\setlength{\epsfxsize}{12.0cm}
\centerline{\mbox{\epsffile{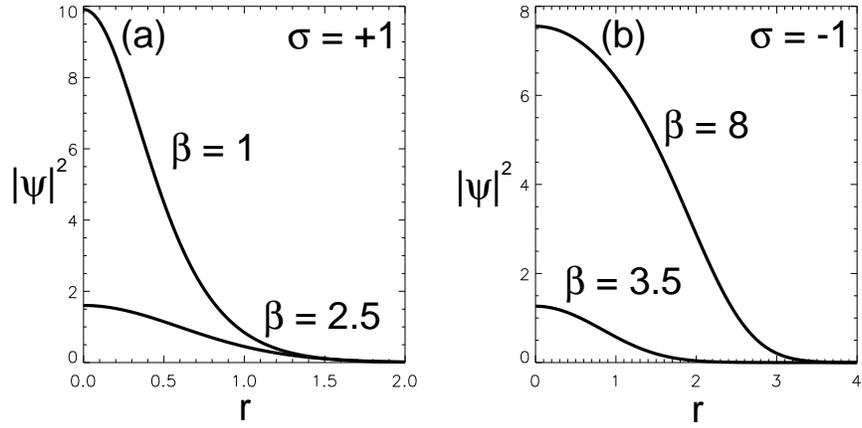}}}
\caption{(a,b) Examples of the condensate density in a 3D trap of radial symmetry for (a) 
attractive ($\sigma = +1$) and (b) repulsive ($\sigma = -1$) interaction, 
respectively.}
\label{fig9}
\end{figure}

\begin{figure}
\setlength{\epsfxsize}{12.0cm}
\centerline{\mbox{\epsffile{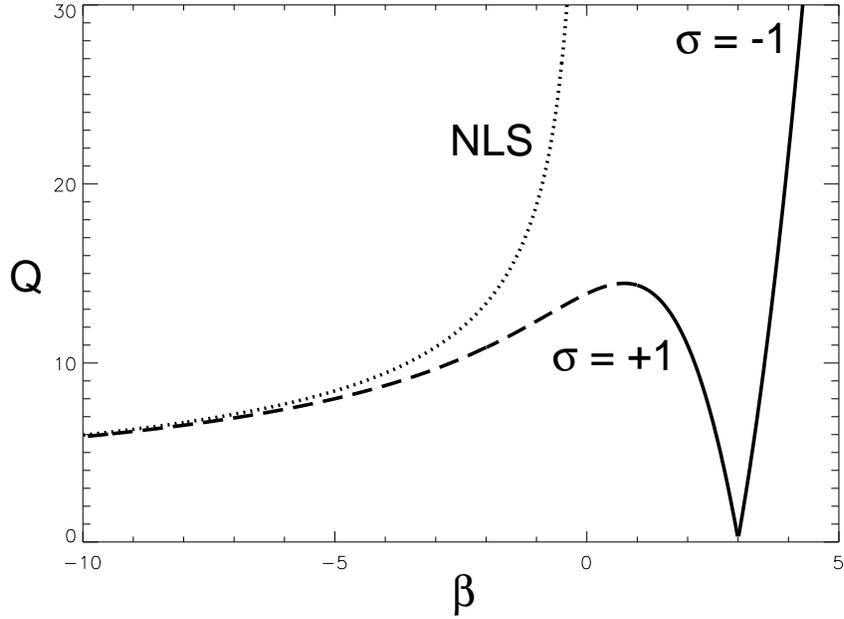}}}
\caption{Invariant $Q$ vs. the parameter $\beta$ for the 
family of the radially symmetric ground-state solutions of the 3D GP equation.}
\label{fig10}
\end{figure}

\end{document}